\RequirePackage{fix-cm}
\documentclass[smallextended]{svjour3}
\smartqed
\usepackage{graphicx}

\begin{document}

\title{Quantization of the Location Stage of Hotelling Model
}

\author{Yuannan Chen \and
        Gan Qin \and
        \nolinebreak An \nolinebreak Min \nolinebreak Wang
}

\institute{Gan Qin \at
              Department of Modern Physics, University of Science and Technology of China, \\
              Hefei, Anhui 230026, China\\
              \email{gqin@ustc.edu.cn}
}

\date{}

\maketitle

\begin{abstract}
We use Li's method \cite{Li} to quantize the location stage of the
Hotelling model. We present the classical model we want to
quantize, and investigate the quantum consequences of the game.
Our results demonstrate that the quantum game give higher profit
for both players, and that with the quantum entanglement parameter
increasing, the quantum benefit over the classical increases too.
Then we extend the model to a more general form, and quantum
advantage keeps unchanged.
\end{abstract}

\section{Introduction}
Entanglement plays an important role in quantum information
theory. Two entangled particles have certain connections even if
they are departed distantly. It is the fundamental physical
principle of the famous quantum teleportation \cite{Bennett}.
Interestingly, entangled particles can be used to play games, e.g,
two players are each distributed with one particle and they make
their movements by operating on his own particle. The idea of this
kind of games had aroused enthusiasm in the so-called quantum game
theory. Since the first paper about quantum game by Meyer
\cite{Meyer}, and shortly afterwards another by Eisert
\cite{J.Eisert}, a series research work about playing games using
quantum objects had been done \cite{series-1,series-2,series-3}.
The early works showed that even if players act non-cooperatively
in quantum games, they could achieve results which could only be
achieved through cooperation in corresponding classical games.
This phenomenon, which could mainly be attributed to entanglement
in particles, had attracted many interests.
\par
In addition to quantizing classical games with discrete
strategies, quantizing the games with continuous strategies is
attracting more attention. In economics, games are usually played
with continuous strategies, such as the quantity, or the price of
the products, or the geographical location of a company. Li \emph{et al}
established a quantization scheme using two single-mode
electromagnetic fields for Cournot duopoly\cite{Li}, a famous
model in micro-economics in which two firms simultaneously choose
their quantities of products. They found that quantum Cournot
duopoly can actually get the result which only can be gained
through cooperation in classical case. Soon the same quantization
scheme were applied to the Bertrand duopoly \cite{Lo} and
Stackberg duopoly \cite{Lo2}. And then the method was extended to
research incomplete information games  \cite{Du,Qin,Lo3} and
asymmetric games \cite{Qin2}.
\par
Hotelling model, which was first presented by Hotelling in 1929,
is another famous model in micro-economics \cite{Hotelling}.
Unlike Cournot duopoly which is basically about choosing
quantities of productions, and Bertrand duopoly which is basically
about choosing the prices of productions, Hotelling model is
initially a spatial duopoly, fundamentally about the choice of
geographical location. But Hotelling himself, had given another
explanation about this model, i.e, locations in this model could
be understood figuratively. In that sense, Hotelling model is
actually about specific character of the products and locations
are used as a measure to show differences between the products on
the market and the ideal products consumers intended to buy.
Besides, the Cournot and Bertrand model basically gave robust
results. We basically need only do small adjustment if we change
the initial settings in these models. But Hotelling model is very
sensitive to initial settings, and slightly changes in its initial
settings can give largely different results. Because of these odd
features, Hotelling model had more than once became a focus of
studies in 20th century and it had developed into many different
forms \cite{Martin}. And it is not yet a closed model \cite{Chen}.
\par
So it would be attractive to try to quantize this model. Although
there are many different versions, Hotelling model is normally a
two-stage game. Firstly the firms choose their locations
simultaneously. Secondly, they choose their prices or quantities
simultaneously. Recently, R.Rahaman quantized the second stage of
this model, by dealing with a certain Hotelling-Smithies model \cite{Rahaman}. In
this paper, we quantize the first stage of the Hotelling model.
\par
In section 2, we recapitulate the original classical Hotelling
model with location choice. In section 3 we discuss the
quantization of a certain kind of Hotelling model, among of which
subsection 3.1 is the introduction of the classical Hotelling
model, and subsection 3.2 is the quantization scheme of the game
and comparison between the classical and the quantum results. In
section 4, we extend the classical game in section 3 into a more
general form and then analyze its quantum results. Section 5 is
the conclusion.
\par

\section{Original Classical Hotelling Model, or the \(D=1\) model }
The original Hotelling model is a spatial duopoly model. As we
have mentioned in section 1, it could be understood literally or
figuratively. In this paper, we would simply understand Hotelling
model literally, although what we present here could also be
explained in the other way. Two firms (which could also be
vendors, restaurants, shops, factories, etc), $A$ and $B$,
providing the same products are located on a one-dimensional
spatial market illustrated as a line segment $CD$ of length $L$ in
Fig. 1, along which there is one consumer per unit length. The
demand of the consumers is totally inelastic and each consumer
would buy one unit product, which is to say the "density demand
function" is $D=1$ along the line segment $CD$, and consequently
the total demand would be $q_D=L$. Firm $A$ is located at distance
$a$ away from the left side and firm $B$ is located at distance
$b$ away from right side. A restriction $0\le a,b\le L/2$ is
supposed here, considering the symmetry of the game. The two firms
sell their product with price $p_1$ and $p_2$ respectively, and
the transportation cost of the consumer is $t$ per unit length. So
the total expense of a consumer located at $s$ ($0\le s\le L$) to
buy one unit product would actually be \(p_1+t\cdot |s-a|\) if he
choose to buy from firm $A$, or \(p_2+t\cdot |s-(L-b)|\) if he
choose to buy from firm $B$. The consumer will compare the total
expense, and chooses the firm with a lower total expense. So there
would be a separation point \(s'\) on the line segment, which
satisfy \(p_1+t\cdot |s'-a|=p_2+t\cdot |L-b-s'|\). Consumers
located in $ s\le s' $ would choose firm $A$ and located in $s\ge
s'$ would choose firm $B$. Thus, the aggregate quantity sold by
each firm is given by $q_1=a+x$, $q_2=b+y$. Here, $x=|s'-a|$,
$y=|L-b-s'|$. For simplicity, we assume the cost of each product
$c=0$. So the profit of the two firms are

\begin{equation}
u_1=p_1\cdot q_1=p_1(a+|s'-a|), u_2=p_2\cdot q_2=p_2(b+|L-b-s'|).
\end{equation}

The two firms are assumed to choose locations $a$, $b$ and prices
$p_1$, $p_2$ to make best profit.

According to Martin \cite{Martin}, this competition model should
be standardized as a two-stage static game: on the first stage,
the two firms simultaneously choose their locations $a$ and $b$;
on the second stage, the two firms simultaneously choose their
prices $p_1$, $p_2$ based on their locations $a$ and $b$. To find
the solution of subgame-perfect equilibrium, we know from game
theory that backward induction can be used here: first we solve
Nash Equilibrium (NE) of the second stage, then we determine the
NE for the first stage \cite{Fudenberg}.
\par

\begin{figure}[ht]
\begin{center}
\includegraphics[height=2cm]{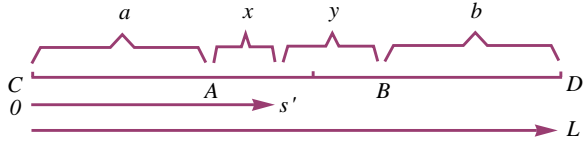}
\caption{Hotelling duopoly on a one-dimensional line market. Firm
$A$ is located at distance $a$ from the left side and firm $B$ is
located at distance $b$ from right side. Consumers located in $
s\le s' $ would choose firm $A$ and located in $s\ge s'$ would
choose firm $B$. $x=|s'-a|$, $y=|L-b-s'|$. }
\end{center}
\end{figure}

\par
Since
\begin{equation}
\left\{
\begin{array}
{l}
p_1+t\cdot x=p_2+t\cdot y \\ a+x+y+b=L
\end{array} \right.,
\end{equation}
\par
we can easily get
\par
\begin{equation}
x=\frac{1}{2}\cdot (L-a-b-\frac{p_1-p_2}{t}),
\end{equation}
\begin{equation}
y=\frac{1}{2}\cdot (L-a-b+\frac{p_1-p_2}{t}).
\end{equation}
\par
Substitute the above results into Eq. (1), we have
\par
\begin{equation}
u_1=\frac{1}{2}\cdot (L-a-b)\cdot p_1-\frac{p_1^2}{2t}+\frac{p_1\cdot p_2}{2t},
\end{equation}
\begin{equation} u_2=\frac{1}{2}\cdot (L-a-b)\cdot p_2-\frac{p_2
^2}{2t}+\frac{p_1\cdot p_2}{2t}.
\end{equation}
\par
On the second stage, \(\partial u_1/\partial p_1=0,\ \partial u_2/\partial p_2=0\). NE would be
\par

\begin{equation}
p_1=t\cdot (L+\frac{a-b}{3}),
\end{equation}
\begin{equation}
p_2=t\cdot (L-\frac{a-b}{3}).
\end{equation}
\par
Thus,
\begin{equation}u_1=\frac{t}{2}\cdot (L+\frac{a-b}{3})^2,\end{equation}
\begin{equation}u_2=\frac{t}{2}\cdot (L-\frac{a-b}{3})^2.\end{equation}
\par
On the first stage, we can easily found out \(\partial
u_1/\partial a>0,\  \partial u_2/\partial b>0\). This means that
firm $A$ or $B$ would improve its profit when it moves nearer
toward each other. With the restriction $0\le a,b\le L/2$, the NE
solution would actually be \(a=L/2,b=L/2.\)

This outcome is not satisfactory. If the firms could cooperate and
both choose $a=b=L/4$, they would make no less profit, with the
transport expense minimized for consumers.
\par

\section{ Simplified version of \(D=1-t\cdot |s-s'|\) Hotelling model}

\subsection{Classical situation}
Here we present the classical model we want to quantize in this
paper, which is a bit different from the original model. In this
model, firstly, we set \(D=1- t\cdot |s-s'|\), where $s'=a$ or $
L-b$ is used to index the location of the firm while $s$ is used to
index consumer's location. This 'density demand function' means
consumer is sensitive to the transport cost, but not sensitive to
the price of the product, which reveals the fact that even the
consumer's demand is inelastic to the price of the products, it
would reduce if the transportation cost becomes high since it is
indeed an extra pay added to the original price. Based on this
'density demand function', we have
\par
\begin{equation}q_1=\int_0^{a+x} (1-t\cdot |s-s'|)ds=a+x-1/2\cdot t\cdot (a^2+x^2),\end{equation}
\begin{equation}q_2=\int_{a+x}^{L} (1-t\cdot |s-s'|)ds=b+y-1/2\cdot t\cdot (b^2+y^2).\end{equation}
\par
Secondly, we assume the price of the product for the two firms is
determined by the supplier and $ p_1=p_2=p_0$. This is often the
case if the two "firms" are in fact two retailers and the retail
price of the products has been fixed to a "unified price" by a
powerful manufacture. As a result, $A$ and $B$ would only compete
on the location choice with fixed price, and the two-stage
Hotelling model was simplified to a one-stage game.
\par
From the first formula of Eq. (2), it is straightly $
x=y=\frac{1}{2}\cdot (L-a-b)$. Then the profit functions turn to
be
\par
\begin{equation}
 u_1=p_0\cdot [a+\frac{1}{2} (L-a-b)-\frac{1}{2}t(a^2+\frac{1}{4}(L-a-b)^2)],
 \end{equation}
\begin{equation}
u_2=p_0\cdot [b+\frac{1}{2}
(L-a-b)-\frac{1}{2}t(b^2+\frac{1}{4}(L-a-b)^2)].
\end{equation}
\par
For this model, noticing that $0\le a,b \le L/2$, three cases of
NE can be obtained as follows:
\par
(1) if $t\in [0,1/L)$, then $\partial u_1/\partial a>0,\partial
u_2/\partial b>0$, NE would be $a=b=L/2$. Accordingly,
$u_1=u_2=p_0 (L/2-L^2 t/8)$.
\par
(2) if $t\in [1/L,2/L]$, the NE require
\begin{equation}\partial u_1/\partial a=p_0\cdot [\frac{1}{2}-\frac{t}{2}\cdot (2a+\frac{a+b-L}{2})]=0,\end{equation}
\begin{equation}\partial u_2/\partial b=p_0\cdot [\frac{1}{2}-\frac{t}{2}\cdot (2b+\frac{a+b-L}{2})]=0,\end{equation}
\par
which lead to $a=b=\frac{2+Lt}{6t}\in [L/3,L/2]$, and consequently
\begin{equation}u_1=u_2=-p_0 t [(\frac{2+Lt}{6t}-\frac{L}{4})^2+\frac{L^2}{16}] +p_0 \frac{L}{2}.\end{equation}
\par
(3) if $t>2/L$, $t$ is out of reasonable range, since $D=1-t\cdot
|s-s'|$ is possibly negative.
\par
It can be easily verified that neither of the profits of case (1)
or (2) are Pareto efficient. Actually if they cooperated and chose
locations as $a=b=L/4$, they could both make higher profits.

\subsection{Quantum situation}

\par
Now we apply Li's method \cite{Li} to quantize the above Hotelling
model. In classical game, the strategies of the two firms is
directly determined by their independent choice of $a$ and $b$. As
a comparison, in quantum game the strategies of the two firms are
determined by their independent choice of two quantum variables
$x_1$ and $x_2$, which have the following relationships with $a$
and $b$:
\par
\begin{equation}
a=x_1 \cosh\gamma +x_2 \sinh\gamma,\label{001}
\end{equation}
\begin{equation}
b=x_2 \cosh\gamma +x_1 \sinh\gamma.\label{002}
\end{equation}
\par
Substituting the above relations into Eq. (13) and (14), we have
\begin{eqnarray}
u_1&=&p_0\cdot [(x_1 \cosh\gamma +x_2 \sinh\gamma)+\frac{t}{2}(L-x_1e^\gamma-x_2e^\gamma)\nonumber\\
& &-\frac{t}{2}(x_1 \cosh\gamma +x_2 \sinh\gamma )^2-\frac{t}{2}(L-x_1e^\gamma-x_2e^\gamma)^2],
\end{eqnarray}
\begin{eqnarray}
u_2&=&p_0\cdot [(x_2 \cosh\gamma +x_1 \sinh\gamma)+\frac{t}{2}(L-x_1e^\gamma-x_2e^\gamma)\nonumber\\
& &-\frac{t}{2}(x_2 \cosh\gamma +x_1 \sinh\gamma )^2-\frac{t}{2}(L-x_1e^\gamma-x_2e^\gamma)^2].
\end{eqnarray}
Similarly to the classical model, the results of the game are
classified by the range of $t$:
\par
(1)If $t\in [0,(1-\tanh\gamma)/L)$, we have $\partial u_i/\partial
x_i>0$, NE would be $a=b=L/2$, and $u_1=u_2=p_0 (L/2-L^2 t/8)$ .
\par
When $\gamma =0$, this interval of $t$ would be $[0,1/L)$, which
corresponds the classical situation.
\par
When $\gamma\to \infty$, this interval of $t$ would reduce to
zero, it means the classical result won't appear at maximal
entanglement.
\par
(2) If $t \in [(1-\tanh\gamma)/L,2/L]$, $\partial u_i/\partial
x_i=0$ has a solution
\par
\begin{equation}
x_1=x_2= \frac{2\cosh\gamma+Lt\cosh\gamma-2\sinh\gamma+Lt\sinh\gamma}{2t(1+2\cosh2\gamma+2\sinh2\gamma)},
\end{equation}
\par
and consequently
\begin{equation}
a=b=\frac{2\cosh\gamma+Lt\cosh\gamma-2\sinh\gamma+Lt\sinh\gamma}{6t\cosh\gamma+2t\sinh\gamma}.
\end{equation}
\par
When $\gamma =0$, the interval of $t$ would be $[1/L,2/L]$, which is the same as the classical situation.
\par
When $\gamma\to \infty$, the interval of $t$ would expand to $(0,2/L]$, the
whole reasonable region except for point $0$, and the NE solution is
$a=b=L/4$.
\par
(3) If $t>2/L$, $t$ is out of reasonable range, as we have
discussed in 3-1.
\par
Now we check out the quantum profit. We are able to calculate the
total quantum profit using the above NE solutions. Here we just
present one firm's profit since the other's is just the same.

\begin{equation}
u_{iq}=\left\{
\begin{array}
{l@{\quad\ \quad}l}
 (\frac{L}{2}-\frac{1}{8} L^2 t) & 0\le t< \frac{1-\tanh \gamma}{L} \\ {\scriptstyle p_0 t[-(\frac{2\cosh \gamma+L t\cosh r-2\sinh \gamma+L t\sinh r}{6 t\cosh \gamma+2 t\sinh \gamma}-\frac{L}{4})^2-\frac{L^2}{16}]+p_0 \frac{L}{2}} &  \frac{1-\tanh \gamma}{L}\le t\le \frac{2}{L}
\end{array} \right.
\end{equation}

In 3-1, we have already got the profit for each firms in the
classical game, which is listed compactly as follows:
\par
\begin{equation}
u_{ic}=\left\{
\begin{array}
{l@{\quad\ \quad}l}
p_0 (\frac{L}{2}-\frac{1}{8} L^2 t) & 0\le t<\frac{1}{L} \\ p_0 t[-(\frac{2+Lt}{6t}-\frac{L}{4})^2-\frac{L^2}{16}]+p_0 \frac{L}{2} &  \frac{1}{L}\le t\le \frac{2}{L}
\end{array} \right.
\end{equation}
\par

Here, the whole interval of $t$, $[0,2/L]$ could be divided into
three regions. In the first region, $0\le t< \frac{1-\tanh
\gamma}{L}$, the profit of quantum game is the same as classical
game, and is independent of $\gamma$. In the second region
$(1-\tanh \gamma)/L\le t<1/L$ and the third region  $1/L\le
t<2/L$, the quantum prifits are different from the classical ones.
\par
In the second region, $(1-\tanh \gamma)/L\le t<1/L$,
\begin{equation}
u_{iq}-u_{ic}=\frac{p_0(2+e^{2\gamma} L t)[-2+(1+e^{2\gamma})L t]}{4(1+ 2e^{2\gamma})^2 t}.
\end{equation}
\par
Figure 2 is the relation curve of quantum profit of one firm and
$\gamma$, and Fig. 3 is the difference between quantum and
classical profit (As a comparison, the first region is also
plotted in the two figures), which indicate that the quantum
profit is more and more exceed the classical one when $\gamma$
increases, and the maximum difference is achieved when $\gamma \to
\infty $.

\par
\begin{figure}[ht]
\begin{center}
\begin{minipage}[ht]{11cm}
\begin{minipage}[ht]{5cm}
\includegraphics[width=5cm]{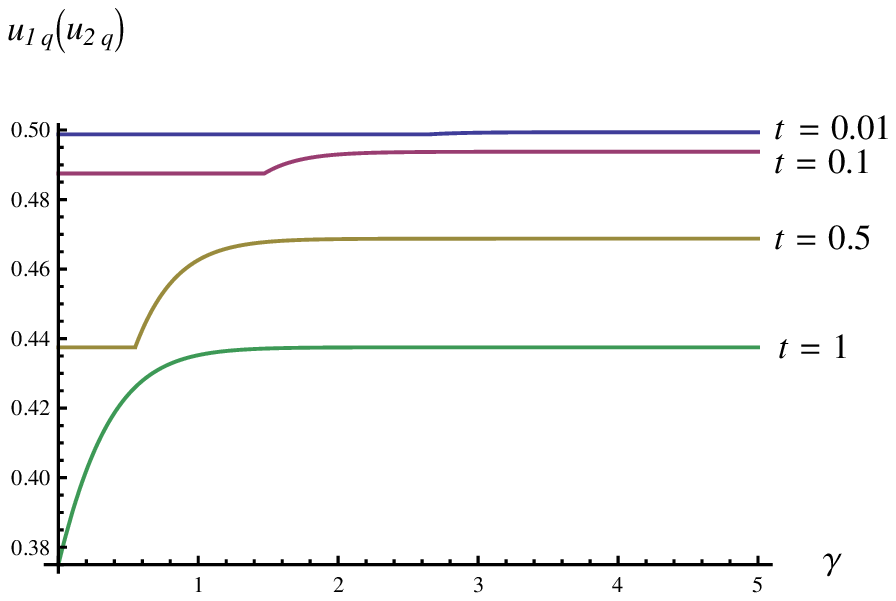}
\caption{Quantum profit for $0\le t<1/L$, which includes the first
region $0\le t< \frac{1-\tanh \gamma}{L}$ (where the curve is
degenerated into horizonal line) and the second region $(1-\tanh
\gamma)/L\le t<1/L$. Here we set $p_0=1,L=1.$}
\end{minipage}
\hfill
\begin{minipage}[ht]{5cm}
\includegraphics[width=5cm]{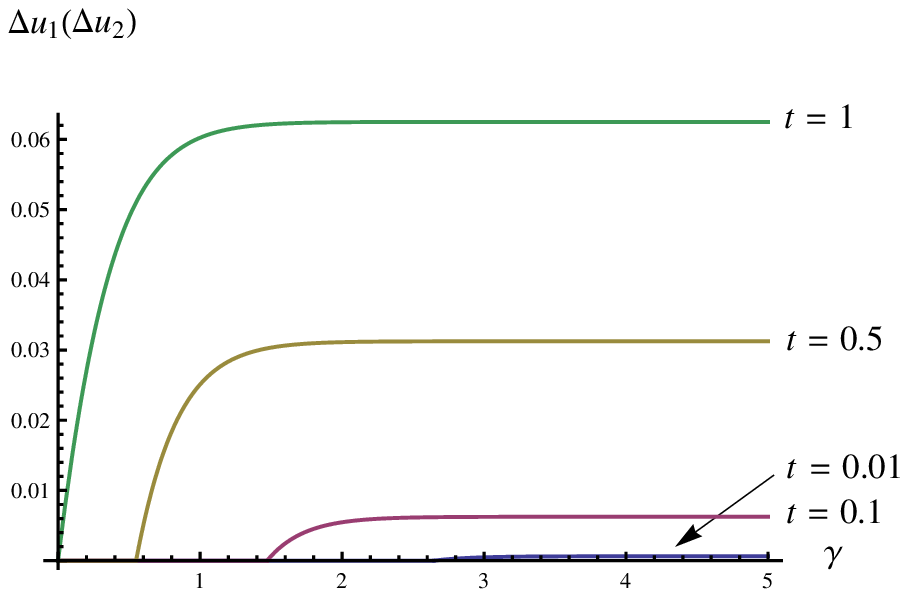}
\caption{Difference between Quantum and Classical situation, for
$0\le t<1/L$, which includes the first region $0\le t<
\frac{1-\tanh \gamma}{L}$ and the second region $(1-\tanh
\gamma)/L\le t<1/L$. Here we set $p_0=1,L=1.$}
\end{minipage}
\end{minipage}
\end{center}
\end{figure}

\par
In the third region, $1/L\le t\le 2/L$, similar analysis leads to
\par
\begin{equation}
u_{iq}-u_{ic}=p_0 t[-\frac{(-4+Lt)^2}{16(t+2e^{2\gamma}t)^2}+(\frac{L}{4}-\frac{2+L t}{6t})^2].
\end{equation}

\par
Figure 4 is the relation curve of quantum profit of one firm and
$\gamma$, and Fig. 5 is the difference between quantum and
classical profit, which indicate the similar pattern as in the
second region, i.e. the quantum advantage becomes more evident
when $\gamma$ increases.

It is remarkable that when $\gamma\to \infty$, the quantum profit
arrives at Pareto optimum. In addition to this, the consumers'
average travel distance is minimized as $L/8$, only a half value
of the classical case.

\par
\begin{figure}[ht]
\begin{center}
\begin{minipage}[ht]{11cm}
\begin{minipage}[ht]{5.0cm}
\includegraphics[width=5cm]{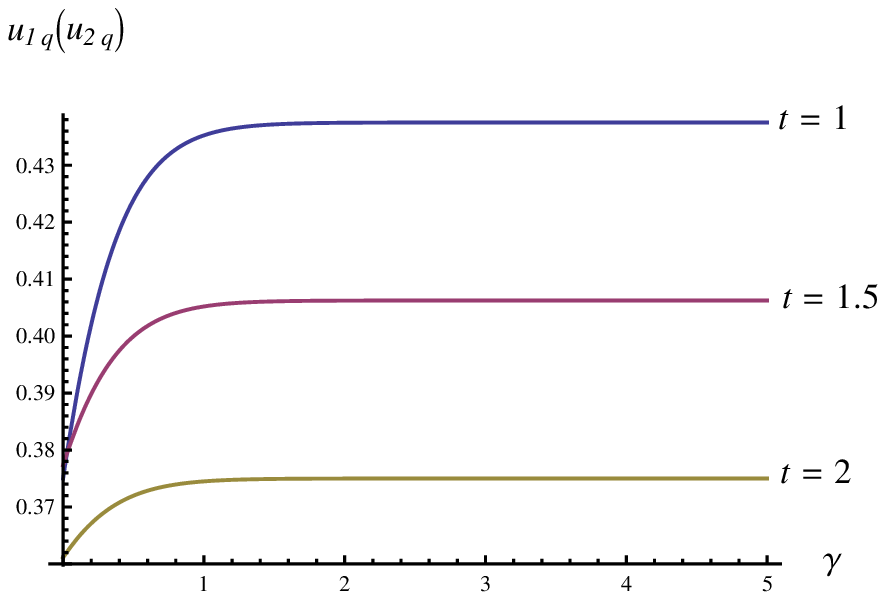}
\caption{Quantum profit for $1/L\le t\le 2/L$. When $\gamma=0$, it
turns back to the classical profit. Here we set $p_0=1,L=1.$}
\end{minipage}
\hfill
\begin{minipage}[ht]{5.0cm}
\includegraphics[width=5cm]{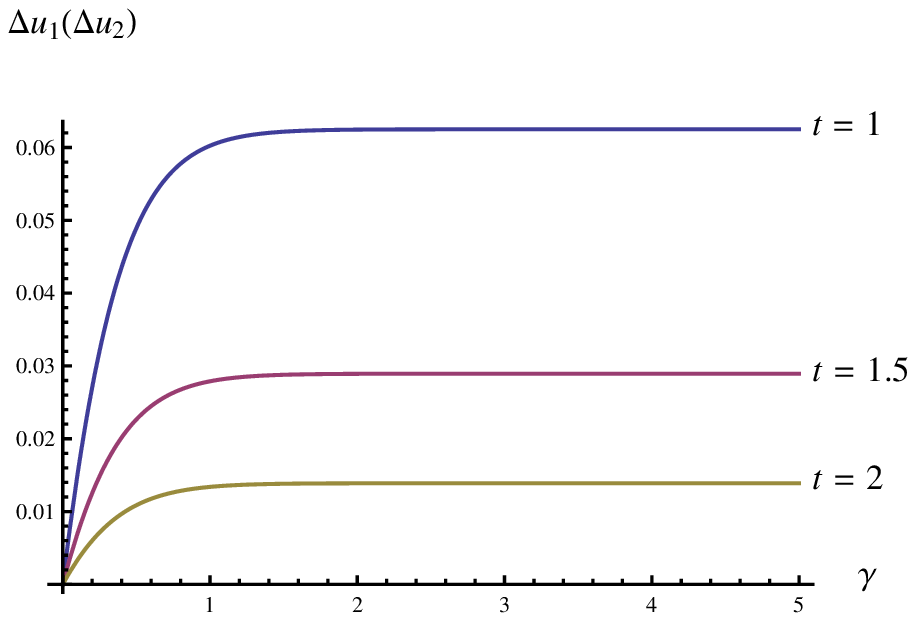}
\caption{Difference between quantum and classical situation for
$1/L\le t\le 2/L$. Here we set $p_0=1,L=1.$}
\end{minipage}
\end{minipage}
\end{center}
\end{figure}
\par
It is not meaningless to discuss the effect of $t$ on the quantum
results. From Fig. 3 and Fig. 5, we can see that the improvement
of profit by quantum scheme increases as $t$ increases up to 1;
then the improvement decreases with the increasing $t$ if $t>1$.
At least such phenomenon shows the complicated coupling effects of
the quantization and the travel cost.

\section{Full version of $D=1-t\cdot|s-s'|$ Hotelling model}

\subsection{Classical situation}

\par
Now we extend the model in section 3 to a more general form. We
abandon the restriction of $p_1=p_2=p_0$, and assume the two firms
are free to choose their prices. This two-stage model could
actually be regarded as a full version of the Hotelling model in
section 3.

In this full version of $D=1-t\cdot|s-s'|$ Hotelling model, we
have
\par
\begin{equation}
u_1=p_1\cdot [a+\frac{1}{2} (\frac{p_2-p_1}{t}+L-a-b)-\frac{1}{2}t(a^2+\frac{1}{4}(\frac{p_2-p_1}{t}+L-a-b)^2)],\label{003}
\end{equation}
\begin{equation}
u_2=p_2\cdot [b+\frac{1}{2} (\frac{p_1-p_2}{t}+L-a-b)-\frac{1}{2}t(b^2+\frac{1}{4}(\frac{p_1-p_2}{t}+L-a-b)^2)].\label{004}
\end{equation}
\par
To solve the equilibrium for this model, we follow the procedure
introduced in section 2: first we solve NE of the second stage,
then we determine the NE for the first stage. For the second
stage, NE requires $\partial u_1/\partial p_1=0$, $\partial
u_2/\partial p_2=0$, and thus we can get the relation of $p_1$,
$p_2$ and $a$, $b$. Then for the first stage, NE requires
$\partial u_1/\partial a=0$, $\partial u_2/\partial b=0$, from
which we can solve the locations, and consequently the prices.

However, the calculation is very complicated. Here we restricted
to the symmetric solutions, i.e. we only look for the solutions
satisfying $a=b, p_1=p_2$. The calculation can be much simplified
with these restrictions.

\par
$\gamma=0$ in Fig. 6 shows the symmetric solution of the location
we found in the classical games. Here we only analysis the
situation when $0\le t\le 1/L$ rather than $0\le t\le 2/L$ when in
section 3. This is due to the fact that the prices of the products
is variables now, and the restriction $0\le t\le 1/L$ is to make
sure that $D$ in this case is positive even if the two firms
choose largely different prices.
\par

\subsection{Quantum situation}

For the quantum situation, we use the same scheme as in subsection
3.2, i.e. quantizing the location strategy while keeping the price
strategy classical. For the quantum strategy indexed by $x_1$ and
$x_2$, we have relationships (\ref{001}), (\ref{002}) presented in
subsection 3.2. The profit function $u_1$ and $u_2$ would be
derived using the relationships (\ref{003}), (\ref{004}). And for
the first stage, we need to investigate the property of $\partial
u_1/\partial x_1$ and $\partial u_2/\partial x_2.$
\par
The quantum ($\gamma =\mbox{sinh} ^{-1} 3/4$, and $\gamma \to
\infty $) and classical ($\gamma=0$) results of the location of
the firms are presented in Fig. 6. And Fig. 7 is the quantum and
classical profit of these cases, which indicates the tendency that
the quantum game gets a better profit for both firms when $\gamma$
increases.
\par
\begin{figure}[ht]
\begin{center}
\begin{minipage}[ht]{11cm}
\begin{minipage}[ht]{5.0cm}
\includegraphics[width=5cm]{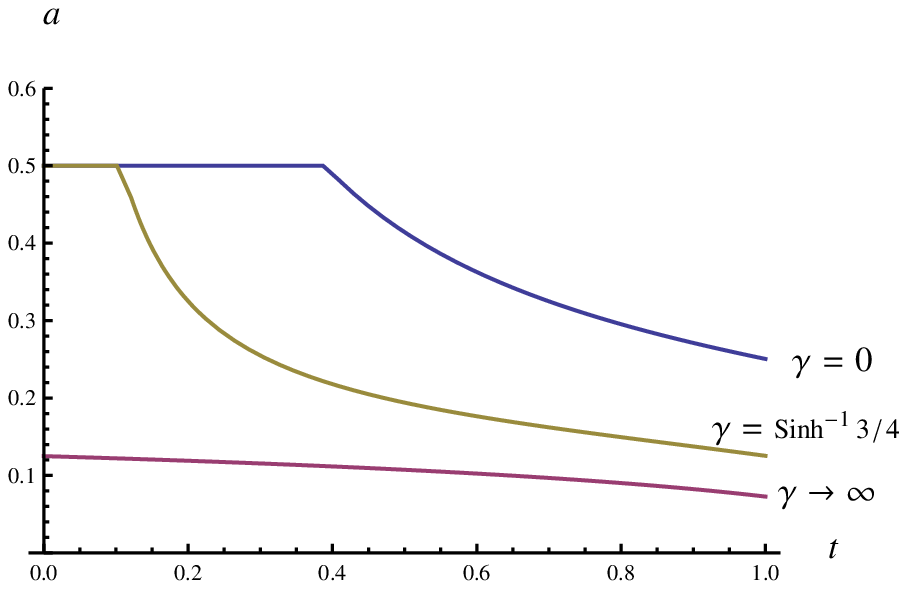}
\caption{Classical ($\gamma=0$) and quantum ($\gamma=\sinh ^{-1}
3/4$ and $\gamma \to \infty$) position for firm $A$ and $B$ as a
function of transport cost $t$. Here we set $L=1$.}
\end{minipage}
\hfill
\begin{minipage}[ht]{5.0cm}
\includegraphics[width=5cm]{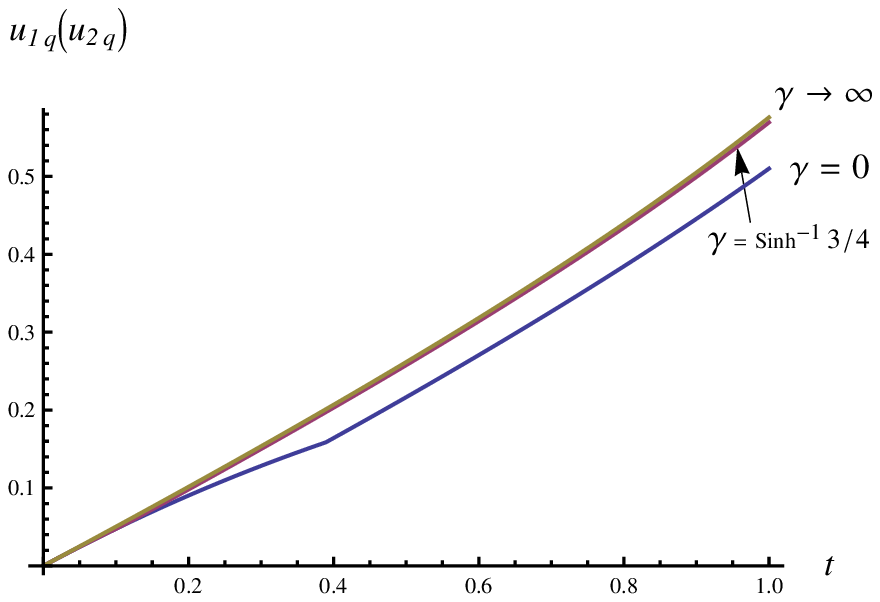}
\caption{Classical ($\gamma=0$) and quantum ($\gamma=\sinh ^{-1}
3/4$ and $\gamma \to \infty$) profit for firm $A$ and $B$ as a
function of transport cost $t$. Here we set $L=1$.}
\end{minipage}
\end{minipage}
\end{center}
\end{figure}
\par
The explicit form of the solutions is generally too complicated to
write down here, just as in the classical game. But luckily for
$\gamma \to \infty $, the solution is simple enough as follows.
\begin{equation}
a=b=\frac{-8+5Lt+\sqrt{64-56Lt+7L^2t^2}}{12t}, \label{location}
\end{equation}
and
{
\begin{equation}
u_1=u_2=\frac{{\scriptstyle[5L^2t^2+Lt(-40+\sqrt{64-56Lt+7L^2t^2})-4(-8+\sqrt{64-56Lt+7L^2t^2})]^2}}{{\scriptstyle 54t(4-Lt+\sqrt{64-56Lt+7L^2t^2})}}.
\label{profit}
\end{equation}
}
This is the best profit the two firms could make using our
quantization scheme.
\par
To better understand the result, let us review the classical game.
In the classical game, if the two firms choose locations in such a
way that $a=b$, and yet compete in the price choice, the profit
could finally be write as
\begin{equation}
u_1=u_2=\frac{t[8a^2t+L(-4-4at+Lt)]^2}{16(2+2at-Lt)}.
\end{equation}
\par
This function is maximized and turned to be (\ref{profit}) when
$a$ and $b$ satisfy (\ref{location}). This is to say, if the two
firms could cooperate in the location stage, they can choose their
locations according to (\ref{location}) to make the best profit
for both of them. Thus the maximal entangled quantum scheme does
help to realize the best profit, which is impossible in the
classical uncooperative game.

\section{Conclusion}
In this paper, we quantized the first stage of Hotelling model,
e.g, the location choice stage, to study the quantum properties of
the game. First we present our version of the model and
investigated the quantum consequences of the game. The quantum
game gave higher profit for both players. And we showed that with
$\gamma$ increasing, the quantum benefit over the classical
increases too. Then we extended the model to a more general form
and we found again, the quantum profit is higher than the
classical one. And there is a common tendency that as $\gamma$
increasing, the quantum advantage becomes more evident.
\par

\begin{acknowledgements}
This work was supported by the National Natural Science Foundation
of China under Grant No. 11375168.
\end{acknowledgements}

\end{document}